\documentclass[prb,showpacs,nofootinbib,twocolumn,amsmath]{revtex4}
\usepackage{hyperref}
\usepackage{bm,amsfonts,mathtools}
\usepackage{graphicx, wasysym}
\usepackage{textcomp}
\usepackage{amssymb,xcolor,latexsym,epsfig}

\newcommand{\be}{\begin{equation}}
\newcommand{\ee}{\end{equation}}

\def\nn{\nonumber}
\def\mL{\mathcal L}
\def\mD{\mathcal D}
\def\mP{\mathcal P}
\def\Fmn{F_{\mu \nu}}

\begin{document}
\date{\today}
\vspace*{0.5cm}

\title{Modified Abelian and SU(2) Wilson theories on a lattice from a noncompact regularization}
                                                                                                                                                                                                                                                                                                                                                                                                                                                                                                                                                                                                                                                                                                                                                                                                                                                                                                                                                                                                                                                                                                    
\author{D. Babusci and F. Palumbo}
\affiliation{
INFN, Laboratori Nazionali di Frascati, 00044 Frascati, Italy} 

\begin{abstract}
Multiflavor gauge theories of matter systems on a three-dimensional lattice have recently been widely investigated especially in connection with a 
possible symmetry enlargement  at a continuous phase transition.  Abelian models were studied both with compact gauge fields and in a mixed 
formulation in which the coupling with matter fields is in compact form while the gauge fields Lagrangian is written in terms of noncompact gauge 
fields, getting quite different results. Such a mixed formulation is not permissible for non-Abelian theories, for which however there exists an entirely 
noncompact  formulation (in which exact gauge invariance is enforced by help of auxiliary fields), which for SU(2)  was shown to yield in the scaling 
window a larger physical volume than Wilson's one. The corresponding U(1) noncompact regularization is derived in the present work. In both 
Abelian and SU(2) cases there is only one auxiliary field that for a large mass has a linear coupling with the other fields and it can be 
integrated out yielding a negative definite correction. The coupling with the auxiliary field might make the inversion of the fermion matrix easier.
\end{abstract}

%\pacs{}
\maketitle

%%%%%%%%%%%%%%%%%%%

\section{Introduction}\label{sec:intro} 
The initial motivation of the present  study was the recent interest in multiflavor gauge theories in three space-time dimensions 
[\onlinecite{Peli1,Bona1,higher-charge,Bona2}], but we think that the results have a wider perspective. Multiflavor gauge theories appear in the investigation 
of condensed matter  [\onlinecite{Klei}] and critical phenomena, especially in connection with a possible symmetry enlargement at a second order phase 
transition [\onlinecite{Sach}]. The relative nonperturbative investigations [\onlinecite{Peli1,Bona1,higher-charge,Bona2}] have been mostly performed by 
lattice Monte Carlo simulations and in several Abelian models the phase diagrams turned out to depend on the way, compact or noncompact, the gauge 
fields are defined on a lattice.

The compact theory was formulated long ago by Wilson [\onlinecite{Wils}] who, in order to enforce gauge invariance in the presence of discrete derivatives, 
replaced the gauge fields in the algebra of a group by the elements of the group, the Wilson variables, which are compact and decompactify in the continuum 
limit.

Later on a new regularization of GL($N_c$) theories on a lattice was defined [\onlinecite{Palu1}], in which  the gauge fields are maintained in the algebra 
of the group, namely they are noncompact, and in order to enforce gauge invariance auxiliary fields are introduced which decouple in the continuum limit. 
Since also the auxiliary fields are noncompact, such a regularization can be called noncompact. For $N_c =2$ the irreducible representation has an 
SU(2) invariance [\onlinecite{Becc1}], while the general definition for SU($N_c >2$) symmetry  has been given in a subsequent work [\onlinecite{Palu2}].  
The construction of such noncompact regularization was done with the idea of being closer to the continuum at finite lattice spacing, an expectation 
supported by a Monte Carlo simulation in four dimensions which showed that in the scaling window it generates for SU(2) a physical space larger than 
the Wilson's one [\onlinecite{DiCa}]. 

In the study of Abelian theories another regularization [\onlinecite{Bona1}] is used, in which  the gauge fields Lagrangian is written as a discretization of the 
continuous one, namely it is noncompact, while the coupling to matter fields is written in terms of  Wilson variables. For this reason, and to distinguish it from 
the entirely noncompact regularization [\onlinecite{Palu1}] that we will use, we will call it half-compact, even though in the literature it is generally referred to as 
noncompact, hoping that this will not generate any confusion. 

Compact and half-compact regularizations in three dimensions give quite different phase diagrams. In the first case the flavor symmetry is broken while 
the gauge fields are in the confined phase; in the second case there are three phases: the Higgs and molecular phases, in which the flavor symmetry is broken 
and the electromagnetic correlations are gapped and ungapped, respectively, and the Coulomb phase in which the flavor symmetry is unbroken and the 
electromagnetic correlations are long range. In view of such striking difference it seems worthwhile to compare them with the entirely noncompact one 
[\onlinecite{Palu1}] that might also introduce further physical behaviors. 

The half-compact  regularization cannot be used for non-Abelian theories, because it breaks the gauge invariance of gauge fields self-interactions and 
therefore a comparison between compact and half-compact was restricted to the Abelian case. Also for non-Abelian theories, however, a comparison 
can be made with  the noncompact regularization, for which there is the additional expectation of a faster convergence to the continuum. 

As a first approach instead of studying  entirely noncompact models, one might investigate the corrections they give to the Wilson Lagrangian  at 
small lattice spacing, in analogy to the modifications obtained by using the renormalization group transformations [\onlinecite{Wils2}] 
or adding some extra terms chosen to cancel the leading lattice artifacts [\onlinecite{Syma}]. These methods will be discussed in Sec. \ref{sec:comp}.
It is the purpose of the present work to derive such corrections.
 
SU(2) and U(1) theories, which are the ones investigated in the quoted multiflavor models [\onlinecite{Peli1,Bona1,higher-charge,Bona2}] have in their 
noncompact form one feature in common: they have only one auxiliary field that in polar variables is a gauge invariant function of the sites. 
For large mass this field has  linear couplings with all the other (gauge and matter) fields. The resulting Lagrangian is  analogous to the gauged 
Nambu--Jona-Lasinio model [\onlinecite{Kogu2}] after linearization: the auxiliary field can then be integrated out leaving a Lagrangian expressed 
in terms of Wilson variables and matter fields only. Of course, in the presence of fermion fields the integration over the auxiliary field has to be performed 
after that on the Grassmann variables, but  in analogy with the gauged Nambu--Jona-Lasinio model, the presence of the auxiliary field might make easier 
the inversion of the fermion matrix (but see later on this point).

The half-compact and noncompact regularizations differ from each other also in another point: only the first one needs a gauge fixing to get rid of 
the pure gauge degrees of freedom that have an infinite volume in the partition function. Some gauge fixings, however, are not compatible with 
some boundary conditions (see for instance Ref. [\onlinecite{Bona3}]),while the choice of boundary conditions may have physical consequences, 
a point  illustrated by a few examples  in Appendix \ref{appA}. In the noncompact regularization [\onlinecite{Palu1}] instead no gauge fixing is 
necessary because a gauge invariant potential damps all the gauge fields, physical and auxiliary, so that all boundary conditions can be handled on 
the same footing. 

Because our modified Lagrangian is valid in any number of dimensions it seems appropriate to review the comparisons made so far in four dimensions.

Abelian theories have been compared in the compact and half-compact forms. It was soon established that the first one does not yield a second 
order phase transition in the bare coupling constant and therefore it cannot define a continuum theory [\onlinecite{Kogu1}]. Several investigations 
have been performed afterwards using the half-compact formulation. After the result that at small coupling the theory is trivial also in this case, 
it was searched whether a continuum theory could be defined at finite coupling [\onlinecite{Azco1}], until it was established that models of Abelian 
gauge fields coupled to scalars [\onlinecite{Baig}] \emph{or} fermions with four flavors in the chiral limit [\onlinecite{Kogu2}] are indeed trivial. Later 
on, however, evidence was reported that in a large-$N$ expansion an ultraviolet fixed point exists in a theory of gauge fields \emph{regularized 
in the continuum and coupled to both fermions and scalars} [\onlinecite{Liti}].

Non-Abelian theories instead have been studied in compact and noncompact form, the latter one however restricted to SU(2) and only in the absence 
of matter fields. First, it was established that the noncompact formulation gives an evidence for confinement [\onlinecite{Poli1}] comparable to that of 
Wilson. This was a prerequisite, because noncompact gauge fields with a simple discretization of the continuum theory  [\onlinecite{Patr}] or a 
nonrenormalizable gauge fixing [\onlinecite{Poli2}] gave a vanishing string tension. Next, a perturbative study of the scaling properties determined 
the renormalization scale parameter and showed that the theory has the same fixed point as  Wilson's one [\onlinecite{Becc2}]. This result was 
confirmed by a calculation [\onlinecite{Bora}] in the Hamiltonian formalism [\onlinecite{Dick}] using L\"uscher's effective Hamiltonian [\onlinecite{Lusc}]. 
Lastly, as already said, the physical volume evaluated in the scaling region [\onlinecite{DiCa}] turned out to be larger than in Wilson's regularization.

In conclusion, in four dimensions there is agreement between half-compact and compact regularizations for Abelian theories coupled to matter fields 
that however turn out to be both trivial (but it should be reexamined in the presence of coupling to both fermions and scalars) and between 
compact and noncompact regularizations of  non-Abelian theories restricted to SU(2) and in the absence of matter fields. 

The paper is organized in the following way. In Sec. \ref{sec:ncU1}  we derive  the definition of gauge transformations  for Abelian fields and the 
relative Lagrangian including interactions with matter fields both in Cartesian and in polar representation, the latter being more convenient for 
a comparison with Wilson's theory.  Such results can be straightforwardly obtained from the general definition for non-Abelian theories 
[\onlinecite{Palu1}] but they have not been spelled out before. In Sec. \ref{sec:imprU1} we derive the U(1) modified Lagrangian. In Sec. 
\ref{sec:ncSU2} we report for the convenience of the reader the known results about models with gauge fields in the irreducible representation 
of SU(2) and construct the corresponding modified Lagrangians. In Sec. V we discuss some previous improvements of the Wilson Lagrangian.

All the above results are valid in four-dimensional Euclidean space. In Sec. \ref{sec:3dim} we briefly indicate how they should be modified to get 
the three-dimensional case. In Sec. \ref{sec:concl} we give our conclusions. We set everywhere $\hbar = c =1$.

%%%%%%%%%

\section{Noncompact U(1) gauge theories on a lattice}\label{sec:ncU1}
We will derive the formulation of noncompact U(1) gauge theories on a lattice in four dimensions from the GL$(N_c) $ ones [\onlinecite{Palu1}], which 
is straightforward but it has not been spelled out so far. We do it for a twofold motivation: the possibility that unlike the compact and half-compact 
regularization it will turn out not trivial (in the absence of coupling with both fermions and scalars), and because it will serve to construct the 
three-dimensional case in Sec. \ref{sec:3dim}.

%%%%%

\subsection{Gauge fields transformations}
The basic ingredient  is  the complex function of the lattice site
\be
D_\mu (x) = \frac1a - g \,[ W_\mu (x) - i\, A_{\mu } (x) ]
\ee
where $a$ is the lattice spacing, $g$ the gauge coupling constant, $W_\mu $ an auxiliary field and  $A_{\mu } (x) $ the gauge field. Under the gauge 
transformation
\be
g(x) = \exp (-i\,\theta(x)) 
\ee
it transforms according to
\be
\label{eq:gtransf}
D'_\mu (x) = g(x) D_\mu (x) g^* (x + \mu), 
\ee
where  $\mu$ in the argument of $g^*(x + \mu) $ represents the unit vector in the $\mu$ direction and $\theta (x)$ is the parameter of the transformation.  
For small $\theta(x)$ the gauge fields transformation can be obtained from those of SU(2) [\onlinecite{Becc1}] 
\begin{align}
A'_\mu (x) &= A_\mu (x)  + \nabla_\mu^+ \theta (x) - a\,W_\mu (x)\,\nabla_\mu^+ \theta (x) \nn \\
W'_\mu (x) &= W_\mu (x) + a\,A_\mu (x)\,\nabla_\mu^+ \theta (x) 
\end{align}
where 
\be
\nabla_\mu^\pm f = \pm\,\frac1a [f (x \pm \mu ) - f(x)]
\ee
are the right/left discreet derivatives of a function $f$.
In the continuum limit $W_\mu$ becomes invariant (and decouples) while $A_\mu$ transforms as the vector potential. For this reason $W_\mu$ was 
called auxiliary, even though at finite lattice spacing the massless gauge field to be identified with the physical vector potential is a linear superposition 
of $A_\mu, W_\mu$ and then the actual auxiliary field is the orthogonal superposition [\onlinecite{Becc1,Palu2}]. 

One can define the gauge field strength 
\be
F_{\mu \nu}(x) = \frac1{i g}\left[D_\mu (x)\,D_\nu (x + \mu) - D_\nu (x)\,D_\mu (x + \nu)\right]
\ee
which transforms according to
\be
\Fmn' (x) = g (x)\,\Fmn (x)\,g^* (x + \mu + \nu) 
\ee
and to first order in the lattice spacing it becomes
\be
\Fmn = \nabla_\mu^+ A_\nu - \nabla_\nu^+ A_\mu \,.
\ee
For a field $\phi$ transforming as
\be
\label{eq:phitra}
\phi' (x) = g (x)\,\phi (x)
\ee
one can define right/left covariant derivatives 
\begin{align}\label{eq:lrcovd}
(\mD^+_{\mu} \phi) (x) &= D_\mu (x)\,\phi(x + \mu) - \frac1a\,\phi(x) \nn\\
(\mD^-_{\mu} \phi) (x)  &=  \frac1a\,\phi(x) - D_{\mu}^* (x - \mu)\,\phi(x - \mu) 
\end{align}
and the symmetric covariant derivative 
\be\label{eq:symd}
(\mD_\mu \phi) (x) = \frac12\,[D_\mu (x)\,\phi(x + \mu) - D_\mu^*(x - \mu) \phi(x - \mu)].
\ee
They all transform according to Eq. \eqref{eq:phitra}. 

%%%%%

\subsection{The noncompact gauge fields Lagrangian}
The Yang-Mills Lagrangian is
\begin{align}
\label{eq:lagYM}
\mL_{\rm YM} (x) &= \frac14\,\sum_{\mu \nu} \Fmn^* (x) \Fmn (x) \\
                            &=  \frac{\beta}2\,\sum_{\mu \nu}   \Big[D^*_\mu (x + \nu) D^*_\nu (x) D_\nu (x) D_\mu (x + \nu) \nn \\
                            &\qquad  - \,D^*_\nu (x + \mu) D^*_\mu (x) D_\nu (x) D_\mu (x + \nu)\Big] \nn \, 
\end{align}
where $\beta = 1/g^2$. One must avoid, however, contributions with $W_\mu \approx 1/a$ for which the covariant derivative does not contain 
an ordinary derivative. To this end one can include a potential which must necessarily be a function of the invariant 
\be
\label{eq:rho_na}
\rho_\mu^2 = D_\mu^* D_\mu = \left(\frac1a - g W_\mu \right)^2 + g^2 A_\mu^2 \,.
\ee
A natural choice for the potential is
\be
\label{eq:Lc}
\mL_{\rm m} =  \frac{\gamma^2}2 \,\sum_\mu \left(\rho_\mu^2 - \frac1{a^2}\right)^2 \,.
\ee
It gives to the auxiliary field $W_\mu$ the mass $4 \gamma^2 / a^2$ which will play the role of an inverse expansion parameter. 
Moreover it damps all the gauge fields, including their zero modes, so making unnecessary a 
gauge fixing\footnote{It must be noticed that the potential $\mL_{\rm m}$ has two minima: 
for $A_\mu = 0$ $$W_\mu = \frac{1 \mp 1}a\,$$ which should both be accounted for. The problem disappears 
using polar fields, as noted for SU(2) in Ref. [\onlinecite{Becc2}].}.

We notice that the theory so defined is strictly local, being a polynomial in the fields. Arbitrary  polynomials in  $\rho^2_\mu$ can be added 
for particular purposes. 

Let us now write down the interactions with matter fields. The Lagrangian of a scalar field $\phi$ transforming as in Eq. \eqref{eq:phitra} is
\be
\mL_\phi = \frac12\,\sum_\mu | {\mathcal D}_\mu \phi |^2 
+ \frac12\,m_\phi^2  | \phi |^2 + {\mathcal V}(| \phi |)
\ee
where ${\mathcal V} (|{\phi} |)$ is a potential and
\be
\left({\mathcal D}_{\mu} \phi^f \right)(x) = \frac12\, \left[ D_\mu (x) \phi^f (x + \mu) - D_\mu^* (x - \mu) \phi^f (x - \mu)\right]
\ee
is the covariant derivative of the scalar field with flavor quantum number $f$. Here and in analogous expressions we adopt  the summation 
convention 
\be
| \phi |^2 = \sum_f  \big (\phi^f \big)^* \phi^f.
\ee 
The Lagrangian of a spinor field $\psi$ transforming as  $\psi'(x) = g(x) \psi(x)$ is
\be
\mL_\psi = i\,\sum_\mu \bar{\psi} \gamma_\mu {\mathcal D}_\mu \psi + m_\psi \bar{\psi} \psi\,. 
\ee
This Lagrangian is usually called naive because it is plagued by the so-called doubling problem [\onlinecite{Mont,Kaplan,PaluNC}] that has 
been tackled in various ways and it  can be handled in the noncompact formulation as indicated in Sec. III. Reference [\onlinecite{PaluNC}]
contains a second order formulation of free chiral fermions on a lattice. Addition of gauge interactions should not give any problem, but a 
difficulty can be encountered to project out in a gauge-invariant way zero modes of auxiliary scalars in the spinor representation.

%%%%%

\subsection{Polar fields}
We derive the explicit expression of the noncompact Abelian Lagrangian with polar fields, which makes the comparison with Wilson's 
theory easier.

The covariant derivative $D_\mu (x) $ in polar variables reads 
\be
D_\mu  =  \rho_\mu \,U_\mu 
\ee
where the polar radius $\rho_\mu$ is the  auxiliary field  (which is obviously gauge invariant) and $U_\mu(x) $ are the Wilson variables 
parametrized in terms of the angular fields $\alpha_\mu (x)$ 
\be
U_\mu = \exp (i\,a\,g\,\alpha_\mu). 
\ee
They are related to the Cartesian fields according to 
\be
\rho_\mu = \sqrt{D_\mu^* D_\mu}\,, \qquad
U_\mu =  \frac1{\rho_\mu}\,\left(\frac1a - g W_\mu + i\,g A_\mu \right)\,.
\ee
With a little abuse we use the same notation $D_\mu $ for the covariant derivative in polar as well as Cartesian fields.

The potential $\mL_{\rm m}$ is already written in Eq.\eqref{eq:Lc} while the noncompact Yang-Mills Lagrangian density 
\eqref{eq:lagYM} in polar form is \\ 
\begin{align}
\mL_{\rm YM} (x) &= \frac{\beta}2\,\sum_{\mu,\nu} \left[\rho_\mu^2 (x + \nu) \rho_\nu^2 (x) - \rho_{P \mu \nu} (x)\right. \nn \\ 
                                         &  \left. \qquad\qquad +\,a^4 \rho_{P \mu \nu} (x)\,U_{P\mu\nu} (x) \right]\,
\end{align}
\be
\label{eq:UPmn}
\rho_{P \mu \nu} (x) = \rho_\nu (x) \rho_\mu (x + \nu) \rho_\mu (x) \rho_\nu (x + \mu)\,,
\ee
and, having taken into account the symmetry of $\rho_{P \mu \nu}$ under $\mu \leftrightarrow \nu$,
\be
U_{P \mu \nu} (x) = \frac1{a^4}\,\left\{1 - \frac12\,\left[U_{\mu \nu} (x) + U_{\nu \mu} (x)\right]\right\}
\ee
with
\be
U_{\mu \nu} (x) = U_\mu^* (x + \nu) U_\nu^* (x) U_\mu (x) U_\nu (x + \mu)\,.
\ee
The above equations can be directly used in numerical calculations. However, to compare with the compact and 
half-compact regularizations and the continuum it is convenient to split the Yang-Mills Lagrangian according to 
\be
\label{eq:splYM}
\mL_{\rm YM}  = \mL_{\rm G}^{\rm W}  + \mL_{\rm int}  + \mL_\rho \,, 
\ee
where
\be
\label{eq:W}
\mL_{\rm G}^{\rm W} =  \frac{\beta}2\,\sum_{\mu,\nu } U_{P\mu \nu}
\ee
is the Wilson Lagrangian,
\be
\mL_{\rm int} (x) = \frac{\beta}2\,\sum_{\mu,\nu}\left[a^4 \rho_{P \mu \nu} (x) - 1\right]\,U_{P \mu\nu} (x) \nn
\ee
is the interaction between the auxiliary field $\rho_\mu$ and the gauge fields, and
\be
\mL_\rho (x) = \frac{\beta}4\,\sum_{\mu,\nu } \left[\rho_\mu (x) \rho_\nu (x + \mu)\,-\,\rho_\nu (x) \rho_\mu (x + \nu)\right]^2\,.
\ee
is the $\rho_\mu$ Lagrangian. Notice that while $\mL_{\rm YM}$ is local in Cartesian coordinates, it has been separated in 
nonlocal polar terms (nonpolynomial when expressed in terms of Cartesian fields). 

The gauge fields partition function in polar coordinates is
\be
T_{\rm G} = \int_0^\infty  d \rho_\mu \rho_\mu^3\int dU_\mu \exp \left[- \sum_x a^4 (\mL_{\rm YM} + \mL_{\rm m} + \mP) \right]
\ee
where $\mP$ is an arbitrary polynomial in $\rho_\mu^2$. We will use also another expression obtained through the following steps. 
We first introduce $\rho_\mu^2$ as integration variable so that neglecting here and in the following a constant factor the partition 
function can be rewritten
\be
T_{\rm G} =  \int_0^\infty  d \rho^2_\mu \int dU_\mu \exp \left(- \sum_x a^4 \mL_{\rm G} \right) 
\ee
where   
\be
\mL_{\rm G} = \mL_{\rm YM} +  \mL_{\rm m} +  \mL_l + \mP
\ee
with
\be
\label{eq:Lagl}
\mL_l = - \frac1{a^4} \sum_\mu \ln (a^2 \rho_\mu^2) \,.
\ee
Next we perform the change of variables
\be\label{eq:rho-t-s}
\rho^2_\mu = \frac1{a^2}\,\left(1+  \frac1{\gamma}\,s_\mu\right)\,,
\ee
so that  the partition function becomes
\be
%\label{eq:TG}
T_{\rm G} = \int_{-\gamma}^\infty\, ds_\mu \int dU_\mu  \exp \left( - \sum_x a^4\mL_{\rm G} \right)
\ee
where now $\mL_{\rm G}$ must be expressed in terms of $s_\mu, U_\mu$. In particular
\be
\mL_{\rm m} =  \frac12\,\sum_\mu  \frac1{a^4}s^2_\mu\,.
\ee

We achieve a simplification if we can choose 
$$
\mP = -\mL_\rho - \mL_l\,.
$$
If we want to stay strictly local, then this is not possible because $\mP$ must be a polynomial in $\rho_{\mu}^2$. However, if we set
\begin{align}
\mP &= - \frac{\beta a^2}{16}\,\sum_{\mu,\nu} \left(\nabla^+_\mu \rho_\nu^2 - \nabla^+_\nu \rho_\mu^2\right)^2  \nn \\
                    & \qquad\qquad - \frac1{2 a^4} \sum_\mu \left(a^4 \rho_\mu^4 - 4 a^2 \rho_\mu^2 + 3 \right)\,,
\end{align}
in the approximation we will use $\mP + \mL_\rho + \mL_l = 0$, and thus
\be
\label{eq:TG-appr}
\mL_{\rm G} \approx \mL_{\rm G}^{\rm W} + \mL_{\rm int} +  \mL_{\rm m}\,.
\ee

A remark is in order here. The Yang-Mills Lagrangian $\mL_{\rm YM}$ is positive definite, but $\mL_{\rm G}  + \mL_{\rm int}$ is not. 
However  $\mL_{\rm m}$ makes $ \mL_{\rm G}$ bounded from below.

%%%%%%%%%%

\section{Modified U(1) Lagrangians from the noncompact one}\label{sec:imprU1}
We firstly observe  that we get  exactly the compact regularization by making the coupling constant $\gamma$ to diverge for vanishing 
lattice spacing [\onlinecite{Palu1}]
\be%\label{eq:lambda}
\gamma = \left(\frac{\lambda}a\,\right)^\eta\,,\qquad \eta > 0
\ee
where $\lambda$ is an arbitrary  parameter with the dimension of a length. The potential term $\mL_{\rm m}$ generates in such a case the 
$\delta$ function $\delta(\rho_\mu (x) - 1/a)$ that eliminates the auxiliary field $\rho_\mu$ from the partition function. 

But, based on the results [\onlinecite{DiCa}], which showed that the noncompact regularization generates for SU(2) a physical space 
larger than the compact one, we  evaluate also for U(1) the contribution arising from the auxiliary field for small lattice spacing, and 
therefore we assume $\eta = 0$.

%%%%%

\subsection{Modified compact Lagrangian}
In order to estimate the effective Lagrangian, we perform the approximation
\be
\label{eq:a2tm}
| s_\mu | \ll \gamma\,, \qquad \rho_\mu \approx \frac1a\,\left(1 + \frac1{2 \gamma}\,s_\mu\right)\,.
\ee
It is easy to check that in this approximation $\mL_\rho +  \mL_l + \mP = 0$, and 
\be
\mL_{\rm int} \approx \frac{\beta}{2 \gamma}\,\sum_{\mu,\nu} (s_\mu + s_\nu)\,U_{P \mu \nu} 
                                     = \frac{\beta}{\gamma}\,\sum_\mu  s_\mu\,\sum_\nu U_{P \mu \nu}\,, 
\ee
so that \eqref{eq:TG-appr} becomes
\be
\mL_{\rm G} \approx  \mL_{\rm G}^{\rm W} + \sum_\mu \left(\frac1{2 a^4}\,s_\mu^2  + 
\frac{\beta}{\gamma} s_\mu\,\sum_\nu U_{P\mu \nu} \right)\,,
\ee
which shows that $\gamma$ (the mass of the $s_\mu$ field, as it can be seen by the rescaling $s_\mu \rightarrow \gamma s_\mu$) plays 
the role of an inverse coupling constant. Therefore 
\be
T_{\rm G} \approx \int_{- \gamma}^\infty  d s_\mu \int dU_\mu \exp \left(- \sum_x a^4 \mL_{\rm G} \right). 
\ee

Now the field $s_\mu$ can be integrated out and for large $\gamma$
\be
T_{\rm G} \approx \int dU_\mu \exp \left(- \sum_x a^4\mL^{\rm mod}_{\rm G} \right)
\ee
with an error of the order of $e^{- \gamma^2}$ and
\be\label{eq:pmn2}
\mL^{\rm mod}_{\rm G}  = \mL_{\rm G}^{\rm W} - \frac{\beta^2 a^4}{2 \gamma^2}\,\sum_\mu
\left(\sum_\nu U_{P \mu \nu} \right)^2\,.
\ee
$\mL^{\rm mod}_{\rm G}$ is the modified gauge fields Lagrangian, which for $\gamma \to \infty$ coincides with that of Wilson while at 
finite $\gamma$ contains a negative definite contribution. For a first investigation one might just evaluate this contribution to the Wilson action 
at the phase transition.

Let us go to scalar fields, whose covariant derivative (omitting the flavor index) is
\begin{align}
\left({\mathcal D}_\mu \phi\right) (x)  &=a\,\rho_\mu (x)\,\left({\mathcal D}^{\rm W}_\mu \phi\right) (x) \nn \\
&\qquad + \frac{a}2\,(\nabla^-_\mu \rho_\mu)\,U^*_\mu (x - \mu)\,\phi(x - \mu)
\end{align}
where
\be
\left({\mathcal D}^{\rm W}_\mu \phi\right) (x) = \frac1{2 a}\,\left[U_{\mu}\,\phi(x + \mu) - U_\mu^* (x - \mu)\,\phi(x - \mu)\right]
\ee
is the Wilson covariant derivative.

The scalar field Lagrangian to leading order in the lattice spacing, after a by parts integration in the action, is therefore 
\be
\mL_\phi \approx  \mL_\phi^{\rm W} + \frac1{\gamma}\, \sum_\mu s_\mu\,\delta\mL^\phi_\mu
\ee
where
\be
\mL_\phi^{\rm W} = \frac12\,\sum_\mu  | \left( {\mathcal D}^{\rm W}_\mu \phi\right) |^2
+  \frac12\,m_\phi^2  |\phi|^2  +{\mathcal V}( |{\phi} |)
 \ee
is the Wilson Lagrangian of the scalar field and 
\be
\delta\mL^\phi_\mu = \frac12 \left\{ | \left({\mathcal D}^{\rm W}_\mu \phi \right) |^2 - \frac14\,\nabla^+_\mu 
\left[\left({\mathcal D}^{\rm W}_\mu \phi\right)^* \phi + \left({\mathcal D}^{\rm W}_\mu \phi\right)\phi^*  \right] \right\}
\ee
is its correction. After integration over the auxiliary field we get the modified Lagrangian 
\be
\mL^{\rm mod} = \mL_{\rm G}^{\rm W} + \mL_\phi^{\rm W} 
- \frac{a^4}{2 \gamma^2}\,\sum_\mu \left(\delta\mL^\phi_\mu + \beta\,\sum_\nu U_{P\mu \nu}\right)^2
\ee
that also contains a negative definite contribution.

The naive spinor field Lagrangian is 
\be
\mL_\psi = i a \rho_\mu \sum_\mu \bar{\psi} \gamma_\mu \mD^{\rm W}_\mu \psi\,+ \,m_\psi \bar{\psi} \psi 
                              +  \frac{i}2 a \nabla^-_\mu  \rho_\mu (\bar{\psi} \gamma_\mu \psi).
\ee
Performing a by part integration in the action and expanding the $\rho$ field, we get to first order in the lattice spacing
\be
\mL_\psi \approx \mL_\psi^{\rm W} + \frac1{\gamma}\, \sum_\mu s_\mu\,\delta \mL^\psi_\mu
\ee
where
\begin{align}
\mL_\psi^{\rm W} &= i\,\sum_\mu \bar{\psi} \gamma_\mu {\mathcal D}^{\rm W}_\mu \psi +  m_\psi \bar{\psi} \psi  \nn\\
\delta \mL^\psi_\mu &= \frac{i}2 \left[ \bar{\psi} \gamma_\mu {\mathcal D}^{\rm W}_\mu \psi - 
                                                   \frac12\,\nabla^+_\mu (\bar{\psi} \gamma_\mu \psi) \right]\,.
\end{align}
In order to see how can we account for the  doubling problem in the present formalism we observe that we can perform the reverse 
transformation from compact to noncompact form by multiplying the Wilson variables $U_\mu$ by $a\rho_\mu$. Therefore, in all the 
cases in which the doubling problem is solved by addition of an appropriate gauge invariant term to the Lagrangian, one must make 
in such a term the replacement $U_\mu\,\to\,a\,\rho_\mu\,U_\mu$ and expand $\rho_\mu$ according to Eq. \eqref{eq:a2tm}.

The final correction to $\mL^{\rm W}_\psi$ has the same form as that of the scalar fields, and it can be handled the same way that 
yields, however, a quartic term in the fermion fields. Usually it is preferable instead to first integrate over the fermion Grassmann variables. 
We notice in this connection that the correction term might make easier the inversion of the fermion matrix in the chiral limit even though, 
at variance with the linearized gauge Nambu--Jona-Lasinio model [\onlinecite{Kogu2}], the help would not come from a masslike term but 
from the fermion current.

%%%%%

\subsection{Modified half-compact Lagrangian}
The half-compact regularization can be obtained from the compact one by expanding the link variables in the pure gauge fields Lagrangian with 
respect to the lattice spacing 
\be
\label{eq:kine}
\sum_{\mu, \nu} U_{P\mu\nu} \approx\,\frac12\,g^2\,\sum_{\mu, \nu} \left(\nabla^+_\mu \alpha_\nu -  \nabla^+_\nu \alpha_\mu\right)^2  
\ee 
which yields the gauge invariant kinetic term of the vector potential. In the coupling to matter fields instead such an expansion would break gauge 
invariance and for this reason the Wilson variables are maintained in the half-compact regularization in the coupling to matter fields.

Again one can look at the difference between the present noncompact regularization and the half-compact one performing an expansion of the 
interaction term to leading order in the lattice spacing. The result is found inserting Eq.\eqref{eq:kine} in Eq.\eqref{eq:pmn2} 
\be
\label{eq:t-alpha}
 - \frac{a^4}{8 \gamma^2}\, \sum_\mu \left[\sum_\nu (\nabla^+_\mu \alpha_\nu 
-  \nabla^+_\nu \alpha_\mu)^2 \right]^2\,,  
\ee
where we took into account that $\beta = 1/g^2$. The Lagrangians of the matter fields get obviously the same corrections as in the noncompact 
regularization.

%%%%%%%%%%

\section{Noncompact SU(2) gauge theories on a lattice}\label{sec:ncSU2}
The formulation of noncompact SU(2) gauge theories has been treated in detail [\onlinecite{Becc1, Becc2}], and most of the results of this section 
about the modification of the compact one can simply be derived from the Abelian case introducing color indices and the relative trace in the proper 
places. Nevertheless we will report more equations than strictly necessary to make this section somewhat self-contained for a reader interested 
in the SU(2) case only.

The basic ingredient for the SU(2) theory is the quaternion
\be
D_\mu (x) = \frac1a - g W_\mu (x)+ i\, g A_{\mu a} (x)  T_a 
\ee
where a is the lattice spacing, $W_\mu $ an auxiliary field, $A_{\mu a} (x) $ the gauge field and $T_a = \sigma_a/2$ the generators of the gauge group, 
$\sigma_a$ being the Pauli matrices. We assume a convention of summation over repeated color indices.  Under the gauge transformations
\be
g(x) = \exp (-i\,\theta_a (x)\,T_a) 
\ee
$D_\mu (x)$ transforms according to
\be
D'_\mu (x) = g(x) D_\mu (x) g^\dag (x + \mu), 
\ee
where $\theta_a(x)$ are the parameters of the transformation.  For small $\theta_a(x)$ the transformations of the gauge fields can be found in 
Ref.[\onlinecite{Becc1}] (where the group generators have a different normalization)
\begin{align}
A'_{\mu a} &= A_{\mu a}+ \left(1 - a W_\mu \right) \nabla_\mu^+ \theta_a - \epsilon_{abc} A_{\mu b} \left(1+ \frac{a}2 \nabla_\mu^+  \right) \theta_c \nn \\
    W'_\mu &= W_\mu + \frac{a}4\,A_{\mu a} \nabla_\mu^+ \theta_a \,.
\end{align}
As in the Abelian case  $W_\mu$ becomes invariant in the continuum (and decouples) while $A_\mu$ transforms as the vector potential, and at finite 
lattice spacing the massless gauge field to be identified with the vector potential is a linear superposition of  $A_{\mu a}, W_\mu$ and then the actual 
auxiliary field is the orthogonal superposition [\onlinecite{Becc1,Palu2}]. 

One can define the gauge field strength 
\be
F_{\mu \nu} (x) = \frac1{i g}\left[D_\mu (x)\,D_{\nu}(x + \mu) - D_\nu (x)\,D_{\mu} (x + \nu)\right]
\ee
which to first order in the lattice spacing becomes
\be
\Fmn = \left(\nabla_\mu^+  A_{\nu a} - \nabla_\nu^+  A_{\mu a} - g\,\epsilon_{a b c} A_{\mu b} A_{\nu c}\right)\,T_a\,.
\ee
It transforms according to
\be
\Fmn' (x) = g (x)\,\Fmn (x)\,g^\dag (x + \mu + \nu) .
\ee
For a field $\phi$ transforming as in Eq. \eqref{eq:phitra} one can define right/left covariant derivatives and the symmetric covariant derivative as in 
Eqs. \eqref{eq:lrcovd}, \eqref{eq:symd}, respectively. They all transform according to Eq. \eqref{eq:phitra}. 

The Yang-Mills Lagrangian is 
\begin{align}
\label{eq:carYM}
\mL_{\rm YM} (x) &= \frac{\beta}8\,\sum_{\mu \nu} \mbox{Tr}  \left[\Fmn^\dag (x) \Fmn (x) \right] \\
                            &=  \frac{\beta}4\,\sum_{\mu \nu}  \mbox{Tr} \left[D^\dag_\mu (x + \nu) D^\dag_\nu (x) D_\nu (x) D_\mu (x + \nu) \right. \nn \\
                            &\qquad\qquad  \left. - \,D^\dag_\nu (x + \mu) D^\dag_\mu (x) D_\nu (x) D_\mu (x + \nu) \right] \nn 
\end{align}
where $\beta = 4/g^2$ and $\mbox{Tr}$ is the trace over colors.

Also with SU(2), in order to avoid  contributions of  $W_\mu \approx 1/a$ for which the covariant derivative does not contain an ordinary derivative, 
one can include in the Lagrangian a potential that must necessarily be a function of the invariant 
\be
\rho_\mu^2 = D_\mu^\dag D_\mu = \left(\frac1a - g W_\mu \right)^2 + \frac14\,g^2 \sum_a A_{\mu a}^2 \,.
\ee
Notice that $D_\mu^\dag D_\mu$ is not a quaternion but it is a real function of the lattice sites  as in U(1), so that the potential has exactly the form 
\eqref{eq:Lc}, which gives to the auxiliary field $W_\mu$ the mass $4 \gamma^2 / a^2$. Moreover it damps all the gauge fields, including their 
zero modes, so making unnecessary a gauge fixing (see Appendix \ref{appA} for details). 

As in the Abelian case an arbitrary polynomial of $\rho^2_\mu $ can be added for particular purposes. In conclusion, the total gauge fields Lagrangian 
is
\be
\mL_{\rm G} = \mL_{\rm YM} + \mL_{\rm m} +  \mP \,.
\label{eq:L_G}  
\ee
We notice that the theory so defined, being a polynomial, is strictly local also for SU(2).

%%%%%

\subsection{Polar coordinates}
In polar coordinates 
 \be
 D_\mu = \rho_\mu U_\mu
 \ee
where
 \be
 U_\mu = \exp( i a \,g \,\alpha_{\mu a} T_a ). 
 \ee
The Yang-Mills Lagrangian density in polar form is
\begin{align}
 \mL_{\rm YM} (x)&= \frac{\beta}4 \sum_{\mu,\nu} \mbox{Tr} \left[\rho^2_\nu (x + \mu) \rho^2_\mu (x) - \rho_{P \mu \nu} (x)  \right. \nn \\
                                         &\qquad\qquad \qquad \left. +\,a^4 \rho_{P \mu \nu} (x)\,U_{P\mu,\nu} (x) \right] 
\end{align}
where
\begin{align}
\rho_{P\mu \nu} (x) &= \rho_\mu (x) \rho_\nu (x + \mu) \rho_\nu (x)  \rho_\mu (x + \nu) \nn\\
    U_{P \mu \nu} (x) &= \frac1{a^4}\,\left\{1 - \frac12\,\left[U_{\mu \nu} (x) + U_{\nu \mu} (x)\right]\right\}
\end{align}
with
\be
U_{\mu \nu} (x) = U_\mu^* (x + \nu) U_\nu^* (x) U_\mu (x) U_\nu (x + \mu)\,.
\ee
are the same as Eq. \eqref{eq:UPmn}, but now the ordering is in $U_{\mu \nu}$.

$\mL_{\rm YM}$ can be decomposed as in Eq.\eqref{eq:splYM} where now Wilson Lagrangian density and interaction of auxiliary and gauge 
fields are
\be
\mL_{\rm G}^{\rm W} = \frac{\beta}4\,\sum_{\mu,\nu } {\rm Tr}\,U_{P\mu \nu}
\ee
 \be
\mL_{\rm int} (x) = \frac{\beta}4\,\sum_{\mu,\nu}\left[a^4 \rho_{P \mu \nu} (x) - 1\right]\,{\rm Tr}\,U_{P \mu\nu} (x)\,, 
\ee
while $\mL_\rho$ remains unchanged. The gauge fields partition function in polar coordinates is
\be
\label{eq:part-rho}
T_{\rm G} = \int_0^\infty  d \rho_\mu \rho_\mu^3\int dU_\mu \exp \left[- \sum_x a^4 (\mL_{\rm YM} + \mL_{\rm m} + 
                    \mP) \right] 
\ee
where $dU_\mu$ is now the Haar measure. After the change of variable \eqref{eq:a2tm}, omitting an unimportant factor, it can be rewritten in 
the form
\be
\label{eq:part-s}
T_{\rm G} =  \int_{-\gamma}^\infty ds_\mu \int dU_\mu \exp \left(- \sum_x a^4 \mL_{\rm G} \right) 
\ee
where  $\mL_{\rm G} $ is the total gauge fields Lagrangian density in polar coordinates  
\be 
\label{eq:gflag}
\mL_{\rm G} = \mL_{\rm G}^{\rm W} + \mL_{\rm int} + \mL_{\rm m} +  \mL_\rho  
+  \mL_l + \mP\,,
\ee
$\mL_l$ being the same as in Abelian case, Eq. \eqref{eq:splYM}.

%%%%%

\subsection{Modified compact SU(2) Lagrangian}
At variance with the Abelian case, the couplings $ \beta, \gamma$ are not independent from each other.  Due to asymptotic freedom, perturbation theory 
is justified and to one loop it gives [\onlinecite{Becc2}]
\be\label{eq:gamma}
\gamma = \gamma_1\,\beta + \gamma_2
\ee
where $ \gamma_1, \gamma_2$ are independent of  $\beta$,  $\gamma_1$ is arbitrary and $\gamma_2$ can be, but it has not been evaluated. This  
result has been confirmed by a calculation in the Hamiltonian formalism [\onlinecite{Bora}] (see Appendix \ref{appB} for details and comparison with 
numerical simulations). In conclusion, we cannot vary $\gamma$ arbitrarily, but in the scaling region it is certainly large so that the field $s_\mu$ is coupled 
linearly to the Wilson variables as in the Abelian case, and it can be integrated out giving the modified gauge fields Lagrangian
\be
\mL_{\rm G} = \mL_{\rm G}^{\rm W} - \frac{\beta^2\,a^4}{8 \gamma^2}\,
\sum_\mu \left[{\rm Tr} \sum_\nu U_{P \mu \nu} (x) \right]^2
\ee
which contains a negative definite contribution. In the deep continuum limit we can use Eq. \eqref{eq:gamma}, getting
\be
\mL_{\rm G} = \mL_{\rm G}^{\rm W} - \frac{a^4}{8 \gamma_1^2}\,
\sum_\mu \left[{\rm Tr} \sum_\nu U_{P \mu \nu} (x) \right]^2\,.
\ee
In the scaling windows of Ref. [\onlinecite{DiCa}], instead, 
\be
0.07\,<\,\frac{\beta^2}{\gamma^2}\,<\,0.44\,.
\ee

Also the evaluation of matter fields Lagrangians proceeds in the same way as in the Abelian case. The covariant derivative of a scalar with flavor 
quantum number $f$ and color quantum number $c$ in the fundamental representation for small lattice spacing reads
\begin{align} 
\left({\mathcal D}_\mu \phi^f \right)^c(x) &= a \rho_\mu (x) \left({\mathcal D}^{\rm W}_\mu \phi^f (x) \right)^c \nn\\
                                                               &+ \frac{a}2\,\nabla_\mu^- \rho_\mu (x)\,\sum_d (U^{\dagger})^{cd}_\mu (x - \mu) \phi^{fd} (x - \mu) \nn\\
\end{align}
where
\begin{align} 
\left({\mathcal D}^{\rm W}_\mu \phi ^f \right)^c(x) &= \frac1{2a}\,\sum_d \left[U^{cd}_\mu (x) \phi^{fd} (x + \mu) \right.\nn \\
                                                                              &\left. \qquad\qquad - (U^\dag)^{cd}_\mu (x - \mu) \phi^{fd} (x - \mu)\right]
\end{align}
is the Wilson covariant derivative. The modified Lagrangian including the interaction with scalars is therefore 
\be
\mL^{\rm mod} = \mL_{\rm G}^{\rm W} + \mL_\phi^{\rm W} - \frac{\beta^2\,a^4}{\gamma^2}\,
\sum_\mu \left(\frac2{\beta}\,\delta\mL^\phi_\mu + \sum_\nu {\mbox {Tr}}\,U_{P\mu \nu}\right)^2 
\ee
while as far as fermions are concerned there is nothing new with respect to the Abelian case.

%%%%%

\section{Comparison with some previous modifications of Wilson action}\label{sec:comp}
Previous improvements of the Wilson action essentially aim at reducing the lattice artifacts. A thorough review of the huge literature on the subject is 
outside the scope of the present work, and we will restrict ourselves to a short discussion of the reduction of lattice effects close to the continuum 
and effects originated by large fluctuations of the plaquette. 

To the  first category belong Wilson's renormalization group [\onlinecite{Wils2}] and Symanzik's approach based on a perturbative expansion with respect 
to the lattice spacing [\onlinecite{Syma}]. Wilson's approach is based on the fact that under a renormalization group transformation the lattice action moves 
toward a fixed point along a definite trajectory, without change of the partition function, that is the same as that of the  original system and describes the 
same long range physics reducing the lattice artifacts [\onlinecite{LusWei}]. Combining analytical with numerical calculations one can get ``perfect" lattice 
actions completely cutoff independent even on coarse grained lattices [\onlinecite{HasNie}]. 

Symanzik's approach is based on the idea of adding extra terms to Wilson's action chosen in a way to cancel the lattice artifacts of $O(a^n)$ to the maximum 
practically possible order [\onlinecite{Syma},\onlinecite{Wei}]. Both methods along with some subsequent developments were reviewed by Niedermayer 
[\onlinecite{Nied}].

Concerning large fluctuations of the plaquette, in order to reduce them two different kinds of actions have been proposed in 
Refs. [\onlinecite{Bane}] and [\onlinecite{Lusc1}], respectively
\be\label{eq:plaq1}
S_p = \beta w + \gamma w^q\,,
\ee
with $w$ as a plaquette action,
\be\label{eq:plaq2}
{\cal L}_{\mu\nu} = \left\{
\begin{array}{l}
\displaystyle{\frac{F^2_{\mu\nu}}{1 - \frac1{\epsilon^2}\,F^2_{\mu\nu}}} \qquad {\mathrm {if}} \qquad | F_{\mu\nu} |\,<\,\epsilon \nn \\
\nn \\
\infty \qquad {\mathrm {otherwise}}.
\end{array} \right.   
\ee
Actually the original purpose of the second one was to formulate a gauge-invariant Abelian chiral theory on the lattice and the cutoff of large 
plaquette fluctuations was instrumental to perform some necessary formal manipulations, but it was later used as a way to reduce plaquette 
fluctuations [\onlinecite{BieFu}].

We notice that for small lattice spacing and plaquette value the last action becomes
\be 
{\cal L}_{\mu\nu} \approx (F_{\mu\nu})^2 \,\left[1 +  \frac1{\epsilon^2}\,F_{\mu\nu}\right]^2
\ee
which is (after the necessary correspondence of parameters is established) identical to ours, Eq. \eqref{eq:pmn2}, but  with opposite sign of the 
correction. Also in the first one the correction has opposite sign with respect to our expression \eqref{eq:pmn2}. 

Our results have been obtained in a spirit completely different from all the above: the heuristic idea is that starting with noncompact fields one is closer 
to the continuum since the beginning as confirmed by Ref. [\onlinecite{DiCa}]. From the technical point of view we remind that our correction has been 
obtained by an expansion in inverse powers of the mass of the auxiliary field and it turns out to be $O (a^4)$. Our results suggest that the actions in 
Eqs. \eqref{eq:plaq1},\eqref{eq:plaq2} should be modified by adding terms correcting for lattice artifacts at small lattice spacing. 

We can observe that the present improvement is very simple and does not need evaluation of other terms than the plaquette. Moreover, we think 
that our improved action might be more convenient than Wilson's one  as a starting point in all the approaches discussed above.

\section{Three-dimensional case}\label{sec:3dim}
In four space-time dimensions the gauge fields have energy dimension 1 and the coupling constant is dimensionless, so that with or without  it all the 
terms in the covariant derivative have the same dimension. In three dimensions instead both gauge fields and gauge coupling constant have energy 
dimension 1/2, but again all the terms of the Wilson variables have the same dimension. The same is true for the angular fields of the Wilson variables
\be
U_\mu = \exp (i\,a\,g\,\alpha_\mu)\,.
\ee

Scalar and fermion fields have dimension 1/2, 1, respectively so that their Lagrangians remain unchanged as well. 

The auxiliary fields $\rho_\mu$ and $s_\mu$ remain of dimension 1, 0 respectively, but the coupling $\gamma$ has dimension -1/2, so that the 
change of variables \eqref{eq:rho-t-s} becomes
\be
\rho^2_\mu = \frac1{a^2}\,\left(1 + \frac{\sqrt{a}}{\gamma}\,s_\mu\right)\,,
\ee
and
\be
\mL_{\rm m} = \frac12\,\sum_\mu \frac1{a^3} s^2_\mu\,.
\ee
For U(1) the modified Lagrangian
\be
\mL^{\rm mod} = \mL_{\rm G}^{\rm W} + \mL_\phi^{\rm W} 
- \frac{a^4}{2 \gamma^2}\,\sum_\mu \left(\delta\mL^\phi_\mu + \beta\,\sum_\nu U_{P\mu \nu}\right)^2\,,
\ee
identical to the corresponding expression in four dimensions \eqref{eq:pmn2} (the dimensions match because $\gamma^2, \beta$ have dimensions -1). 
All the changes in the other expressions are obvious.

%%%%%

\section{Conclusions}\label{sec:concl}
In recent works U(1) models in three dimensions were compared in compact [\onlinecite{Bona1}] and half-compact form [\onlinecite{higher-charge}] 
getting quite different phase diagrams, while SU(2) models (for which a half-compact formulation does not exist)  were studied only in compact form.  
It seems therefore interesting to extend the comparison to an entirely noncompact lattice formulation that was defined  for general non-Abelian 
theories. Such regularization showed  for SU(2) a faster convergence to the continuum, but  the price to be paid  is the introduction of one auxiliary 
gauge invariant field. 

For a first investigation one can estimate the corrections to the Wilson Lagrangian for large mass of the auxiliary field, which can therefore be integrated out 
yielding a Lagrangian expressed in terms of compact gauge fields and  matter fields only. For a further simplification one can evaluate the contribution 
of the correction only at the phase transition.

The modified Lagrangian is quartic in the matter fields, that is impractical  in the presence of fermions, in which case it may be convenient to retain 
the auxiliary field, whose Lagrangian is similar to that of the gauged Nambu--Jona-Lasinio model [\onlinecite{Kogu2}] after linearization. As in the 
latter the linear coupling with the auxiliary field might help the inversion of the fermion matrix.

The present result can be extended to QCD, but in such a case there are color auxiliary fields and $D^\dag_\mu D_\mu$ is no longer a gauge invariant 
function of the sites but it has an expression involving the generators of the group. We hope to be able to tackle this problem in a future work.

%%%%%

\section{Acknowledgments}
We thank M.-P. Lombardo for most interesting conversation, correspondence and suggestion of references.

%%%%%%%%%%%%%%%%%%%

%\appendix

\section{Appendix A: Boundary conditions}\label{appA}
Because pure gauge fields are not damped in the partition function, half-compact gauge theories need a gauge fixing, but as already noticed some 
gauge fixings are not compatible with some boundary conditions [\onlinecite{Bona3}]. The difficulties arising from periodic boundary conditions for 
quantization in a cubic box, for instance, can be avoided giving the gauge field a mass vanishing in the continuum limit [\onlinecite{Azco3}], or
assuming antiperiodic boundary conditions [\onlinecite{Bona3}] or, at the price of nonlocality, with a complete axial gauge fixing [\onlinecite{Palu5}]. 
 
Boundary conditions however cannot be chosen on the basis of mere calculational convenience, because they may have physical effects; quite 
generally different boundary conditions favor or disfavor different phases [\onlinecite{Camp}] or select different physical systems as illustrated 
by the following examples.
 
L\"uscher constructed an effective action for continuous QCD in four dimensions with periodic boundary conditions in terms of zero-momentum modes 
(gauge fields depending on time but constant in space) integrating out all the other modes [\onlinecite{Lusc}]. The mass spectrum so obtained differs 
from that derived [\onlinecite{Gonz}] assuming the twisted boundary conditions (which forbid zero-momentum modes) firstly studied by 't Hooft 
[\onlinecite{'tHo}]. Indeed the latter boundary conditions force fluxes of magnetic and electric fields in the quantization box, defining a physical system 
different from the one studied by L\"uscher.

Further examples are provided by QED. At strong coupling in the compact version with periodic boundary conditions electric charges are confined, 
while with appropriate quasiperiodic boundary conditions one gets a Coulombic potential [\onlinecite{Palu3}]. Moreover, with periodic boundary 
conditions zero-momentum modes may contribute to radiative corrections [\onlinecite{Palu4}].

%%%%%%%%%%%%%%%%%%%

\section{Appendix B: Perturbative calculations versus numerical simulations}\label{appB}
The are two sources of difficulties in a direct  comparisons among  perturbative calculations and numerical simulations: 
\begin{enumerate} 
\item [1.] The choice of the partition function \eqref{eq:part-rho} or \eqref{eq:part-s}. \\
\item [2.] The choice of the arbitrary polynomial  $\mathcal {P}$ in the gauge fields Lagrangian \eqref{eq:gflag}.
\end{enumerate}

Concerning point 1, in both perturbative calculations [\onlinecite{Becc2, Bora}] the  partition function \eqref{eq:part-s} was adopted with the 
approximation $\gamma \gg 1$ in the lower bound of integration over $s_\mu$. Therefore to be compared with the quoted perturbative results the 
numerical simulations should have $\gamma \gg 1$ in the scaling window. 

Now, in the simulation [\onlinecite{DiCa}] the partition function \eqref{eq:part-rho} was evaluated with the integral over $s_\mu$ extending from  
$- \infty$, which implies the same approximation $\gamma \gg 1$ as in the perturbative calculations, while in the simulation [\onlinecite{Poli1}] the 
partition function\eqref{eq:part-rho} was adopted, with the integral over $\rho_\mu$ extending from zero to $+ \infty$, namely without any approximation 
as far as this integral is concerned.
 
The scaling window results to be $\gamma \approx 4$ in Ref. [\onlinecite{Poli1}] while in Ref. [\onlinecite{DiCa}] there are two scaling regions with 
$\gamma > 8$ and $ 6 > \gamma > 10$.

Concerning point 2, the arbitrary polynomial has been chosen $\mP = - \mL_\rho$ in the perturbative calculation [\onlinecite{Bora}] 
and in the simulation [\onlinecite{Poli1}], while it  has been chosen $\mP = - \mL_\rho - \mL_l$ in the perturbative calculation 
[\onlinecite{Becc2}] and in the simulation [\onlinecite{DiCa}].

In conclusion, a direct comparison among all these results is not straightforward.

%%%%%%%%%%%%%%%%%%%

\end{document}